 \documentclass[12pt]{article}
 \usepackage{amsfonts,amssymb}
 \renewcommand{\labelenumi}{(\alph{enumi})}
 
\newtheorem{Def}{Definition}[section]

\newtheorem{The}[Def]{Theorem}

\newtheorem{Exa}[Def]{Example}

 \newcommand{\End}{\nonumber\\ }
 \newcommand\Mbox[1]{ \mbox{\rm{#1}}}
 \newcommand\Mboxq[1]{\quad \mbox{\rm{#1}}\quad}
 \newcommand{\Textfrac}[2]{{\textstyle{\frac{#1}{#2}}}}
 \newcommand\Texthalf{\Textfrac{1}{2}}
 \newcommand{\Deff}[1]{\textsl{#1}}
 \newcommand{\Lrb}[1]{\left( #1 \right)}
  \newcommand{\Set}[1]{\left\{ #1 \right\}}
 \newcommand{\One}{1}
 \newcommand{\Sum}[2]{\sum_{#1}^{#2}}
 \newcommand{\Ker}{\mathrm{Ker}}
 \newcommand{\Ima}{\mathrm{Im}}
 \newcommand{\Coh}{\mathrm{H}}
 \newcommand{\Ipa}[2]{\left\langle #1, #2 \right\rangle}
  \newcommand{\Comm}[2]{\left[ #1 , #2 \right]}
  \newcommand{\Liebracc}[2]{\left[ #1 , #2 \right]}
 \newcommand{\Pb}[2]{\left\{ #1 , #2 \right\}}
 \newcommand{\Exp}[1]{\exp\left(#1\right)}
 \newcommand{\Tx}[1]{{\textstyle{{#1}}}}
 \newcommand\Real{ {\mathbb R} }
 \newcommand{\Df}{ {\mathrm{d}}}
 \newcommand{\Dbd}[1]{\frac{\Df}{\Df #1} }
 \newcommand{\Pbd}[1]{\Tx{\frac{\partial}{\partial #1} }}
 \newcommand{\Pbdt}[2]{\frac{\partial #1}{\partial #2}}
 \newcommand{\Ldiff}[1]{\mathcal{L}_{#1}}
 \newcommand{\Dc}{\mathcal{D}}%{ {\rm D}}
  \newcommand{\Al}{\alpha}
  \newcommand{\Be}{\beta}
  \newcommand{\De}{\delta}
 \newcommand{\Ep}{\epsilon}
 \newcommand{\Et}{\eta}
 \newcommand{\Th}{\theta }
 \newcommand{\Io}{\iota }
 \newcommand{\La}{\lambda}
 \newcommand{\Rh}{\rho}
 \newcommand{\Si}{\sigma}
 \newcommand{\Om}{\omega}
 \newcommand{\LA}{\Lambda}
 \newcommand{\Expower}[1]{\LA^{#1}}
 \newcommand{\Resmap}[1]{\big|_{#1}}
 \newcommand{\alphi}{\renewcommand{\labelenumi}{(\alph{enumi})}}
 \newcommand{\Lg}{\mathfrak{g}}
 \newcommand{\Lh}{\mathfrak{h}}
 \newcommand{\Lk}{\mathfrak{k}}
 \newcommand{\Lgd}{\mathfrak{g}^*}
 \newcommand{\Lhd}{\mathfrak{h}^*}
 \newcommand{\Adr}[1]{\mathrm{Ad}_{#1}}
 \newcommand{\Scon}[3]{C_{#1#2}^{#3}}
 \newcommand{\Cotan}[1]{T^*{#1}}
 \newcommand{\Sdpgh}{G\rtimes H}
 \newcommand{\Itrans}[2]{R_{#1}{}^{#2}}
 \newcommand{\Ghrep}[2]{U_{#1}{}^{#2}}
 \newcommand{\Reduc}[2]{M_{#1}{}^{#2}}
 \newcommand{\Lagr}{\mathfrak{L}}
 \newcommand{\Dlagr}[1]{\frac{\delta \Lagr}{\delta {x}^{#1}}}%{\delta \dot{x}^{#1}}}
 \newcommand{\Pman}{\mathcal{N}}
 \newcommand{\Pmane}{\mathcal{N}'}
 \newcommand{\Mmap}{T}%{\mathbf{J}}
 \newcommand{\Curr}{J}%{\mathbf{J}}
 \newcommand{\Tmap}{\mathbf{\phi}}
 \newcommand{\Fpman}{\Fsmo (\Pman)}
 \newcommand{\Spman}{\mathcal{SN}}
 \newcommand{\Uvar}{w}
 \newcommand{\Brst}{\textsc{brst}}%{{\rm{\textsc{BRST}}}}
 \newcommand{\Bv}{\textsc{bv}}
 \newcommand{\Brstop}{Q}%\OM
 \newcommand{\BrstH}{Q_H}
 \newcommand{\BrstG}{Q_G}
 \newcommand{\Diffone}{\De}
 \newcommand{\Difftwo}{\Df}
 \newcommand{\Difftot}{\mathrm{D}}
 \newcommand{\Elh}{u}
 \newcommand{\Man}{\mathcal{M}}
 \newcommand{\Cotanm}{T^*\Man}
 \newcommand{\Fsmo}{\mathcal{F}}
 \newcommand{\Uone}{U(1)}
 \newcommand{\Kof}{\tilde{X}}
 \newcommand{\Hil}{\mathcal{H}}
 \newcommand{\Gtil}{\tilde{G}}
 \newcommand{\Lgt}{\tilde{\mathfrak{g}}}
  \newcommand{\Pvec}[1]{\underline{#1}}
  \newcommand{\Jtil}{\tilde{\Mmap}}
 \newcommand{\Jsi}{\Mmap_{\Si}{}}
  \newcommand{\Transg}[1]{\mathrm{Tr}(#1)}
 \newcommand{\Symuv}{\Om'}
  \newcommand{\Wstar}{W^*}
  \newcommand{\Hstar}{H^*}
  \newcommand{\DiffE}{D_E}
  \newcommand{\DiffF}{D_F}
  \newcommand{\DiffA}{D_A}
  \newcommand{\END}{{\mathrm{End}}}
 \newcommand{\Rps}[2]{#1/\!\!/#2}
\begin{document}
\begin{center}
 {\large Equivariant \Brst{} quantization and reducible symmetries}\\
 \ \\
   Alice Rogers\\
  \ \\
  Department of Mathematics      \\
  King's College             \\
  Strand, London  WC2R 2LS
  \
  \\
  alice.rogers@kcl.ac.uk%
 \end{center}

 \vskip0.2in
 \begin{center}
 April 2007
 \end{center}
 \vskip0.2in

 \begin{abstract}
Working from first principles, quantization of a class of  Hamiltonian systems with reducible
symmetry is carried out by  constructing first the appropriate reduced phase space and then
the \Brst{} cohomology. The constraints of this system correspond to a first class set for a
group $G$ and a second class set for a subgroup $H$. The
\Brst{} operator constructed is equivariant with respect to  $H$.
Using algebraic techniques analogous to those of equivariant de Rham theory, the \Brst{}
operator is shown to correspond to that obtained by \Bv{} quantization of a class of systems
with reducible symmetry. The `ghosts for ghosts' correspond to the even degree two generators
in the Cartan model of equivariant cohomology. As an example of the methods developed, a
topological model is described whose
\Brst{} quantization relates to the equivariant cohomology of a manifold under a circle
action.
\end{abstract}
 \section{Introduction}
 In this paper we derive from first principles a \Brst{} procedure for
quantization of certain symmetric Hamiltonian systems for which the constraints do not form a
closed algebra, because the symmetries are what is known as reducible. It is shown that the
resulting
\Brst{} operator compares to the standard
\Brst{} operator for the related irreducible symmetry in the same way that in equivariant de
Rham cohomology the equivariant derivative compares to the standard exterior derivative. The
auxiliary even generators which occur in this equivariant cohomology  correspond to the
`ghosts for ghosts' in the \Bv{} quantization of reducible symmetries first formulated by
Batalin and Vilkovisky \cite{BatVil,BatVil81}.

The underlying motivation for this work is a desire to understand the functional integral
methods which have proved so powerful in the quantization of theories with symmetry. An
example is constructed showing that the procedure leads to a full path integral quantization
scheme complete with a quantum gauge-fixing procedure, so that quantum calculations are
possible.

In a standard symmetric Hamiltonian system there is a set of `first class' constraints
$T_a, a=1, \dots,m$ on the phase space of the system which are closed under Poisson bracket:
 \[
  \Pb{T_a}{T_b} = \Scon abc T_c \, .
 \]
The constraints are a reflection of the symmetry of the system under some group
$G$, and the true phase space of the system is the quotient of the constrained surface by the
group action. (A more intrinsic, group theoretic formulation is described below in
Section~\ref{secRPS}.) In the simplest situation the coefficients $\Scon abc$ are constants
and the constraint algebra is a finite dimensional Lie algebra, but it is often the case that
the coefficients $\Scon abc$ are more general functions on the phase space, although the
system may still possess symmetry related to a finite-dimensional Lie group, as explained in
Appendix~\ref{appLA}. There are however symmetric systems for which the constraint algebra
does not close, with some of the constraints being second class. The purpose of this paper is
to show that for a class of such systems there is an analysis in terms of a symmetry group
$G$ acting equivariantly with respect to a subgroup
$H$, and to derive the corresponding \Brst{} quantization scheme.  The result is that the true
phase space of the system is obtained in two stages, first reducing the phase space by the
action of $H$ and then by $G$. (Such two stage reduction is described extensively by Marsden,
Misiolek,   Ortega,   Perlmutter and Ratiu in \cite{MarMisPerRat98,MarMisOrtPerRat04}.)

The \Brst{} operator obtained is equivalent to that used in
\Bv{} quantization of first order reducible systems  \cite{BatVil,BatVil81,FisHenStaTei}, but
the derivation is more fundamental, using the algebraic features of the constraints to
construct the appropriate reduced phase space. The even, ghost-number two, fields of the
\Bv{} formalism correspond to the degree two generators of the dual of the Lie algebra of
$H$ in the Weil model of $H$-equivariant cohomology. The relation between
equivariant cohomology and \Brst{} quantization of certain topological
theories has been pointed out by a number of authors, including Kalkman, Chemla and Kalkman
and Ouvry, Stora and van Baal
 \cite{Kalkma,CheKal,OuvStoVan}. In this paper we take these ideas further and give general
arguments based on canonical quantization and the necessary modification of the
Marsden-Weinstein reduction process
 \cite{MarWei} for the open constrained systems studied, obtaining both a more general and a more
fundamental explanation of this connection.

The structure of the paper is that in Section~\ref{secRPS} we first describe the symplectic
geometry of a standard constrained system with closed algebra, including the moment map and
the Marsden-Weinstein reduction procedure \cite{MarWei} leading to the reduced phase space of
the system obtained after gauge redundancy has been removed. We then describe the more general
systems considered in this paper, and the corresponding reduced phase space. (In
Appendix~\ref{appLA}, which relates to systems with closed constraint algebras as well as the
more general systems considered in this paper, we explain that the reduction process -- and
hence the
\Brst{} procedure -- are also applicable in the setting of a more
general class of group action, in which an infinite-dimensional group acts on the phase space,
with a related action of the finite dimensional group $G$ which is only local;  this allows
for the possibility of `structure functions' and apparent variations in the constraint
algebra.) In Section~\ref{secBRST} the standard Hamiltonian \Brst{} method, due to Henneaux
 \cite{Hennea}, Kostant and Sternberg  \cite{KosSte} and Stasheff  \cite{Stashe88a} is reviewed,
while in Section~\ref{secMBRST} the equivariant \Brst{} operator appropriate for the more
general constrained system described in section \ref{secRPS} is constructed. Using methods
adapted from equivariant de Rham theory, various different but equivalent models of the
\Brst{} cohomology are presented. The class of reducible symmetries which leads to the
constrained systems studied in this paper is described in Section~\ref{secRSC}. In the final
section a specific model exhibiting these structures is described. The \Brst{} operator is
that constructed by Kalkman
 \cite{Kalkma} and by Chemla and Kalkman  \cite{CheKal}, but our derivation is from a simple
classical action. The model itself is equivalent to the supersymmetric model introduced by
Witten  \cite{Witten82} and used by Witten and others to obtain powerful equivariant
localization techniques, as may be seen for instance in references
 \cite{Atiyah85,BirBla,Bismut85,Blau93,CorMoo,JefKir}.
\section{The reduced phase space of a partly open constraint
algebra}\label{secRPS}
In this section we will consider the classical dynamics of a Hamiltonian system defined
on a $2n$-dimensional symplectic manifold $\Pman$ on which an $m$-dimensional Lie group
$G$ acts freely and symplectically on the left, with $m \leq n$. To establish notation,
$gy$ will denote the image of the point $y$ in $\Pman$ under the left action of $g$ in
$G$, and for each $\xi$ in the Lie algebra $\Lg$ of $G$, $\Pvec{\xi}$ will denote the
corresponding vector field on $\Pman$. The group action is required to be Hamiltonian, so that
there exists a map (referred to as the \Deff{constraint map})
 $\Mmap:\Lg \to \Fpman$, $\xi \mapsto \Mmap_{\xi}$ (where $\Fpman$ denotes the space of smooth
 functions on $\Pman$) which satisfies the conditions
 \begin{eqnarray}\label{MMAPeq}
   \Ldiff{\Pvec{\xi}}f &=& \Pb{\Mmap_{\xi}}{f} \End
   \Mmap_{\xi}(gy) &=& \Mmap_{\Adr{g}\xi}(y)
 \end{eqnarray}
 for all $f$ in $\Fpman$, $y$ in $\Pman$ and $g$ in $G$  \cite{GuiSte84}. (Here $\Pb{}{}$
 denotes Poisson bracket with respect to the symplectic form on $\Pman$.)
This is the standard set up for a constrained Hamiltonian system: the constraint functions are
the $m$ functions $\Mmap_a \cong \Mmap_{\xi_a}$ corresponding to a basis
 $  \Set{\xi_a|a=1, \dots, m}$
of $\Lg$, and the constraint submanifold $C$ is the subset of $\Pman$ consisting of points
$y$ such that $\Mmap_{a}(y)=0$ for $a=1, \dots, m$. More intrinsically, $C$ is the set $\Tmap^{-1}(0)$
where
 $\Tmap:\Pman \to \Lgd$ is the  the moment map, which is the transpose of the constraint map
 $\Mmap$ and thus defined by
 \begin{equation}
    \Ipa{\Tmap(y)}{\xi} = \Mmap_{\xi}(y)
 \end{equation}
for all $y$ in $\Pman$ and $\xi$ in $\Lg$. By the properties (\ref{MMAPeq}) of the  map
$\Mmap$,
$C$ is invariant under the action of $G$;  the Marsden-Weinstein reduction theorem  \cite{MarWei} states that
the quotient manifold $C/G$ is a symplectic manifold with symplectic form $\nu$ determined
uniquely by the condition
 $\pi^*\nu = \Io^* \Om$, where $\Om$ is the symplectic form on $\Pman$,
 $\Io:C \to \Pman$ is inclusion and $\pi:C \to C/G$ is the canonical
projection.  This result can be proved using theorem 25.2 of  \cite{GuiSte84}, which
establishes that in certain circumstances if $\Om$ is a closed form on a manifold $X$ then the
set of vector fields $\Pvec{\xi}$ on $X$ which satisfy
 \begin{equation}
    \Io_{\Pvec{\xi}} \Om = 0
 \end{equation}
forms an integrable distribution, and the corresponding foliation is fibrating; further, if
$\Rh:X \to \Man$ is the fibration, there is a symplectic form
$\nu$ on $\Man$ uniquely determined by
$\Om=\Rh^*(\nu)$. The symplectic manifold obtained by this two stage reduction process is referred to as the
reduced phase space of the system and will be denoted $\Rps{\Pman}{G}$.  It is the true phase
space of the system; however it is in general a rather complicated space, even when $\Pman$ is
simple, and may not admit a polarisation as required in geometric quantization to determine
the position/momentum split. The
\Brst{} approach, which is described in Section~\ref{secBRST}, is a cohomological formulation
which readily allows a quantization scheme which can be used for path integral quantization,
provided this was the case for the unconstrained phase space. (The situation described so far
is a rather idealised, oversimplified one, most real physical systems involve quantum field
theory rather than quantum mechanics, and the group action can be identified locally rather
than globally. There is a discussion of these matters in appendix \ref{appLA}, in the rest of
this section we will continue to work in the idealised framework so that the key ideas can be
presented in a simple manner.)

A modification of this reduced phase space structure will now be described, which will lead in
section \ref{secMBRST} to a construction which aims to provide the appropriate modification of
the \Brst{} procedure for a system with what is called reducible symmetry. This concept was
first introduced and studied by Batalin and Vilkovisky \cite{BatVil,BatVil81} in the
Lagrangian formalism, and is discussed in section \ref{secRSC}. A key feature is that there is
only a partial symmetry of the system under the action of a Lie group. In the Hamiltonian
framework described above, this means that only some of the constraints are satisfied.  An
important idea in the current paper is that the missing constraints can be incorporated  into
the formalism by introducing new variables. At this stage it is not clear whether the
procedures described are applicable to all first order-reducible systems, further work is
required here, but it is clear that the method described does provide a more fundamental
account of the \Brst{} operator in the reducible case, clarifying the role of `ghosts for
ghosts', and relating them to the even generators of
$S(\Lgd)$ which appear in the de Rham models of equivariant cohomology.

In the canonical setting the ingredients of the systems to be considered again include a
Hamiltonian action of a Lie group
$G$ on a symplectic manifold $\Pman$. Additionally,
$G$ has an Abelian subgroup
$H$ with a particular property, and the  constraints  take the form
 \begin{equation}
    \Mmap_{a} - \Ipa{v}{\xi_a} = 0 \, ,
 \end{equation}
 where $v$ is an arbitrary element of $\Lhd$, the dual of the Lie algebra $\Lh$ of $H$,
 and $\Mmap$ is the constraint map as before. (This
 form of the constraints is related to the Lagrangian approach to reducible symmetries in Section \ref{secRSC}.)
   Constraints of this form are possible if we extend the phase space $\Pman$ by taking the
Cartesian product with $\Cotan{H}$.

The property required of
$H$, which among other things ensures that
$\Lhd$ can be uniquely identified as a subspace of $\Lgd$, is that there is a subspace
$\Lk$ of $\Lg$ such that
 \begin{equation}\label{HPROPeq}
  \Lh \oplus \Lk = \Lg  \qquad
   \Mbox{and} \qquad       \Liebracc{\Lh}{\Lk} \subset\Lk\, .
 \end{equation}
(An example is when $G$ is semisimple and $\Lh$ is a Cartan subalgebra of $\Lg$.) This
property means that $\Lhd$ can be identified as the subspace of $\Lgd$ whose elements $\Elh$
satisfy $\Ipa{\Elh}{\xi}=0$ for all  $\xi$ in $\Lk$.   It will be useful to use a basis
 \begin{equation}\label{HKBASISeq}
 \Set{\xi_{\Al}, \xi_r | \Al= 1, \dots, l, \, r = 1+l, \dots, m}
 \end{equation}
of $\Lg$, with $\Set{\xi_{\Al}|\Al= 1, \dots, l}$ a basis of $\Lh$
($l$ being the dimension of $H$) and $\Set{ \xi_r | r = 1, \dots,
m-l}$ a basis of $\Lk$, and use the notational convention that Greek
letters are used as indices for elements of bases of $\Lh$ while
Latin indices from the second half of the alphabet are used for
$\Lk$ and from the first half for $\Lg$ as a whole.
 The only structure constants $\Scon abc$ with respect to this basis which are non zero are then
those of the form $\Scon{\Al}rs, \Scon rst$ and $\Scon rs{\Al}$.

The extended phase space
$\Pmane=\Pman\times\Cotan{H}$ has symplectic form
 \begin{equation}
    \Symuv = \Om +   \Df v_{\Al}\wedge\Df \Uvar^{\Al},
 \end{equation}
where the variables $v_{\Al}$  are interpreted as coordinates on the cotangent space at each
point of
$H$, which is identified with $\Lhd$, and $\Uvar^{\Al}$ are coordinates on $H$.
(The summation convention for repeated indices is used except when explicitly stated to the
contrary.) In terms of these coordinates the constraints take the form
 \begin{equation}\label{MCONSeq}
 v_{\Al}-\Mmap_{\Al} = 0 \, , \Al=1, \dots l \qquad \Mbox{and} \qquad
    \Mmap_{r}  = 0 , r=1+l, \dots m \, ,
 \end{equation}
where $v_{\Al}=\Ipa{v}{\xi_{\Al}}$ is a coordinate on the extended phase space $\Pmane$. These
constraints  do not in general form a closed algebra under Poisson bracket, instead one has a
partly second class system, neither do they correspond to a $G$ action on $\Pman$ or $\Pmane$.

These problems stem from the fact that we have so far overlooked the fact that the variables
$v_{\Al}, \Al=1, \dots, l$ have canonical conjugates $\Uvar^{\Al}$ which are also constrained.
By taking this into account it  will  be shown that the reduced phase space for the system is
in fact
$\Rps{\Pmane}{(\Sdpgh)}$ where
$\Sdpgh$ denotes the semi direct product of
$G$ and $H$ corresponding to the action of $H$ on $G$ by inverse
conjugation and the action of  $\Sdpgh$ on $\Pmane$ corresponds to the constraint map whose
components with respect to bases
$\Set{\xi_{\Al},\xi_r}$ of $\Lg$ and $\Set{\La_{\Al},\Al=1,\dots,l}$ of $\Lh$ are
respectively
 \begin{equation}
    \Mmap_{\Al}, \Mmap_{r}\Mboxq{and}v_{\Al}-\Mmap_{\Al}.
 \end{equation}
(The Lie algebra of $\Sdpgh$ corresponds to that of the direct product $G \times H$ with
additional non-zero brackets
$\Liebracc{\La_{\Al}}{\xi_r}=-\Scon{\Al}rs \xi_s$.)

This reduced phase space will be obtained from the physical constraints (\ref{MCONSeq}) if
$v_{\Al},\Al=1,\dots,l$ are regarded as  dynamical variables, with canonically conjugate momenta
$\Uvar^{\Al}$.  Since the original Lagrangian from which the constraints have been derived
will not have had any dependence on the time derivative of
$v_{\Al}$, additional constraints $\Uvar^{\Al}=0, \Al=1, \dots, l$
must also be satisfied. Thus we have a system with $m+l$ primary
constraints
$\Mmap_r, v_{\Al}-\Mmap_{\Al},\Uvar^{\Al}, r=1+l,\dots,m,\Al=1,\dots l$.

There are also secondary constraints: since the Hamiltonian of the system is independent of
$\Uvar^{\Al}$, the equation of motion for $v_{\Al}$ is simply $\dot{v}_{\Al}=0$, so that
$v_{\Al}$ is constant.  We choose $v_{\Al}=0$, which then gives us $l$ secondary constraints.
The full set of constraints is thus
 \begin{equation}
    \Set{\Mmap_{r},\Mmap_{\Al}-v_{\Al},v_{\Al},\Uvar^{\Al}, r=1+l,\dots,m,\Al=1,\dots,l} \,.
 \end{equation}
From the Poisson bracket
 $\Pb{v_{\Al}-\Mmap_{\Al}}{\Uvar^{\Be}}=-\De^{\Be}_{\Al}$ we see that the constraints
 $\Uvar^{\Al},v_{\Al}-\Mmap_{\Al}, \Al=1,\dots,l$ form a second class set which
reduces the extended phase space $\Pmane$ to $\Rps{\Pmane}{H}$ under the action of $H$
generated by $v_{\Al}-\Mmap_{\Al}$, which we can represent explicitly as the subspace of
$\Pmane$ where $u^{\Al}=0$ and $v_{\Al}-\Mmap_{\Al}=0$ for $\Al=1,\dots l$ with symplectic
form whose Poisson brackets are given by the Dirac bracket
 \begin{equation}\label{DBRACeq}
    \Pb{f}{g}_D = \Pb{f}{g}-\Pb{f}{v_{\Al}-\Mmap_{\Al}}\Pb{g}{\Uvar^{\Al}}
    +\Pb{g}{v_{\Al}-\Mmap_{\Al}}\Pb{f}{\Uvar^{\Al}} \,.
 \end{equation}

The remaining constraints $\Mmap_r, r=1+l,\dots,m$ and $v_{\Al},{\Al}=1,\dots,l$ form a first
class set on this reduced space (where we can in fact replace $v_{\Al}$ by $\Mmap_{\Al}$), and
the corresponding reduction process then reduces this space further.  This two stage reduction
can be effected all at once by the action of $\Sdpgh$ as indicated above.  A very
comprehensive study of two stage reduction, with applications in a number of classical
contexts, has been made by Marsden, Misiolek, Ortega, Perlmutter and Ratiu in
\cite{MarMisPerRat98} and \cite{MarMisOrtPerRat04}.

In the following section the \Brst{} quantization procedure for a closed constraint algebra
will be described, while in Section~\ref{secMBRST} it will be shown how this construction may
be modified to take into account the reduced phase space of the kind just described,
corresponding to an action of $\Sdpgh$ on an extended phase space.
\section{The {\rm\Brst{} }procedure for a closed constraint
 algebra}\label{secBRST}
In this section we review the \Brst{} procedure for the standard reduced phase space
corresponding to a closed constraint algebra. The reduced phase space is the space
$\Rps{\Pman}{G}=C/G$ constructed in  Section~\ref{secRPS}, with $C=\Tmap^{-1}(0)$.  The formulation of
\Brst{} cohomology in the canonical setting was first given by Henneaux  \cite{Hennea} and by McMullan
 \cite{McMull}, providing a powerful development of the BFV construction  of the vacuum
generating functional of a gauge theory  \cite{FraVil1,BatVil,FraFra,BatFra,FraVil2}.  The
\Brst{} construction was expressed in a more abstract mathematical setting by Kostant and
Sternberg  \cite{KosSte} and by Stasheff  \cite{Stashe88a,Stashe97}.

The idea is to construct a \Brst{} operator $\Brstop$ whose zero degree cohomology theory
agrees with the space of smooth functions $\Fsmo(\Rps{\Pman}{G})$ on the reduced phase space,
and also to construct a super phase space so that the \Brst{} operator
$\Brstop$ is implemented by Poisson bracket. The exposition here largely follows  \cite{KosSte}.
The operator is constructed in two stages.  First, we define a superderivation
 \[
    \Diffone:\Expower{q}(\Lg) \otimes \Fpman \mapsto \Expower{q-1}(\Lg) \otimes \Fpman
 \]
by its action on generators:
\begin{equation}\label{DIFFONEeq}
    \Diffone(\pi \otimes \One) = \One \otimes \Mmap_{\pi},
    \quad\quad
    \Diffone(\One \otimes f) = 0
\end{equation}
where $\pi \in \Lg$ and $f\in \Fpman$. It follows immediately that
 $\Diffone^2=0$. (This is the Koszul complex, as seems first to have been observed by McMullan.)
 Also $\Ker^0 \, \Diffone = \Fpman $ while
 $\Ima^0 \, \Diffone =\Diffone(\Lg) \Fpman $.  Now $\Diffone(\Lg) \Fpman$
is the ideal of $\Fpman$ consisting of functions which vanish on the constraint surface $C$,
and the space of smooth functions on $C$ can be identified with $\Fpman $ modulo this ideal.
Thus $\Fsmo(C)\cong \Coh^0(\Diffone)$, and the first part of the construction of the \Brst{}
operator has been achieved.

To complete the construction, suppose that $K$ is a $\Lg$ module,
and define the operator
 \[
 \Difftwo : K \to \Lgd \otimes K
 \]
by setting $\Ipa{\Difftwo\,k}{\pi} = \pi\, k $ for all $\pi$ in
$\Lg$ and $k$ in $K$.  This operator can be extended to become a
superderivation
\[
 \Difftwo:\Expower{p}(\Lgd) \otimes K \to \Expower{p+1}(\Lgd) \otimes K
\]
by defining $\Difftwo \Et$ for $\Et$ in $\Lgd$ to be the exterior derivative of $\Et$ regarded
as a left invariant one form on $G$. (This, as observed by Stasheff, is the the standard
Chevalley-Eilenberg differential for the Lie algebra cohomology of $G$.) Using the fact that
$\Difftwo$ on
$\Lgd$ is the transpose of the bracket on
$\Lg$, it can be shown that $\Difftwo^2=0$.  Also, it follows from the definition that
$\Ker^0\, \Difftwo$ is equal to the set $K^{\Lg}$ of $\Lg$ invariants in $K$ while
 $\Ima^0 \, \Difftwo$ is zero.  Thus $\Coh^0 \Difftwo$ is equal to $K^{\Lg}$.

 If we now set $K = \Expower{} (\Lg) \otimes \Fpman$,  with the $\Lg$ action on $K$
 defined by
 \begin{eqnarray}
  \lefteqn{
       \xi(\pi_1\wedge \dots \wedge\pi_q \otimes f) =} \End
   && \Sum{r=1}{q} \,\pi_1 \wedge \dots \wedge \pi_{r-1}\wedge\Liebracc{\xi}{\pi_r}
    \wedge\pi_{r+1} \wedge \dots\wedge\pi_q \otimes f  \End
    && \qquad \qquad + \qquad \,\pi_1 \wedge \dots \wedge \pi_q \otimes
     \Pb{\Mmap_{\xi}}{f} \, ,
  \end{eqnarray}
 then the $\Lg$ action commutes with the action of $\Diffone$ on
 $K$, so that $\Diffone$ and $\Difftwo$ commute and $\Difftwo$ is well defined on
the $\Diffone$ cohomology groups of $K$.  Thus
 $\Coh^0(\Coh^0(\Expower{}(\Lgd) \otimes \Expower{} (\Lg) \otimes \Fpman))$
is well defined and equal to the $\Lg$ invariant elements of
$\Fsmo(C)$, and thus to $\Fsmo(\Rps{\Pman}{G})$.

The properties of the two differentials may be summarised in the
diagram
\[ \begin{array}{ccc}
   \Expower{p}(\Lgd) \otimes \Expower{q}(\Lg) \otimes \Fpman &
    \stackrel{\Diffone}{\to} & \Expower{p}(\Lgd) \otimes \Expower{q-1}(\Lg) \otimes \Fpman  \\
    & & \\
   \Difftwo \downarrow &  &  \\
    & & \\
   \Expower{p+1}(\Lgd) \otimes \Expower{q}(\Lg) \otimes \Fpman  &  &
 \end{array}
\]
and we have a double complex
 \[
  \Difftot:\Expower{}(\Lgd) \otimes \Expower{}(\Lg) \otimes \Fpman \to
   \Expower{}(\Lgd) \otimes \Expower{}(\Lg) \otimes \Fpman \, .
 \]
with $\Difftot=\Difftwo+(-1)^p \Diffone$. If we define the total degree of an element of
 $\Expower{p}(\Lgd) \otimes \Expower{q}(\Lg) \otimes \Fpman$
to be $\,p-q$, then $\Difftot$ raises degree by one. Under certain technical assumptions
 \cite{KosSte} $\Coh^0\Difftot$ is equal to
 $\Coh^0(\Coh^0(\Expower{}\Lgd \otimes \Expower{} \Lg \otimes \Fpman))$,
so that we have constructed a complex whose zero cohomology is equal to
$\Fsmo(\Rps{\Pman}{G})$, in other words to the observables on the true phase space of the
system.

If we now construct the $(2n,2m)$-dimensional symplectic supermanifold
 \[\Spman=\Pman \times \Real^{0,2m}\]
 with symplectic form
 $\omega + \Df \pi_a \wedge \Df \Et^a$ where
  $\pi_a, \Et^a,a=1, \dots, m$ are natural coordinates on the $\Real^{0,2m}$
  factor, then $\Fsmo(\Spman)\cong \Expower{}(\Lgd) \otimes \Expower{}(\Lg) \otimes \Fpman$
and $\Difftot$ can be realised by taking Poisson bracket with the
function
 \[
   \Brstop = \eta^a \Mmap_a - \Texthalf \Scon{a}{b}{c} \Et^a \Et^b \pi_c \,.
 \]
 As a result the Poisson brackets with respect to the symplectic form
 $\omega + \Df \pi_a \wedge \Df \Et^a$ close on the zero cohomology of
$\Difftot$
 and correspond to the Poisson brackets on the reduced phase space.
Quantization of this system is straightforward, given a quantization on the original
unconstrained phase space $\Pman$. The Hilbert space of states is taken to be
 $\Hil \otimes \Fsmo(\Real^{0,m})$, where $\Hil$ is the space of states for $\Pman$.  A typical element
is $f_{a_1 \dots a_p}\Et^{a^1} \dots \Et^{a^p}$ where each $f_{a_1 \dots a_p}$ is in $\Hil$
and
$\Et^a,a=1,\dots,m$ are natural coordinates on $\Real^{0,m}$.  (If the
$G$ action is  local, in the manner described in Appendix~\ref{appLA},
then the super phase space and the space of states may be twisted
products rather than simply cartesian products, and the operators $\Diffone$ and $\Difftwo$
defined locally but in a globally consistent manner.)

Observables on the super phase space will be operators of the form
 \[
 A_{a_1 \dots _p}^{b_1\dots b_q}\Et^{a^1} \dots \Et^{a^p}\pi_{b_1}\dots\pi_{b_q}
 \]
 where each
$A_{a_1 \dots a_p}^{b_1\dots b_q}$ is an observable on $\Pman$. The
observables $\Et^a$, which are known as `ghosts', are represented on
states as multiplication operators, while the ghost momenta $\pi_b$
are represented by
 \begin{equation}
    \pi_b = -i \Pbd{\Et^b} \, .
 \end{equation}

An obvious but important consequence of this scheme is that the quantized \Brst{} operator
$\Brstop$ has square zero. We can thus implement the \Brst{} cohomology at the quantum level
by defining physical observables to be observables which commute with $\Brstop$, the quantized
\Brst{} operator, modulo observables which are themselves commutators with $\Brstop$.
Physical states are then  defined to be states annihilated by $\Brstop$ modulo those in the
image of $\Brstop$.  (Further aspects of \Brst{} quantum dynamics, including the gauge fixing
necessary in the path integral approach, are described in  \cite{HenTei}, \cite{GFBFVQ} and
 \cite{GFEC}.)
\section{The modified procedure for a class of open constraint algebras}\label{secMBRST}
In this section we construct the analogue of the \Brst{} procedure for the case where the
constraints and reduced phase space are those of Section~\ref{secRPS}, corresponding to an
$l$-dimensional subgroup $H$ of our $m$-dimensional symmetry group $G$.
The operator will be expressed in a form that allows gauge-fixing and path integral
quantization as in \cite{Witten82} and \cite{GFBFVQ}.

Proceeding directly with the $\Sdpgh$ action on $\Pmane=\Pman\times T^*(H)$ with constraint
map whose components are the constraint functions
$\Mmap_{\Al},\Mmap_r$ and $v_{\Al}-\Mmap_{\Al}=0$,  we obtain the
\Brst{} operator
 \begin{equation}
    \Brstop =  \Et^a \Mmap_a +\Th(v_{\Al}-\Mmap_{\Al})
    - \Texthalf \Et^a \Et^b \Scon abc \pi_c +  \Th^{\Al}\Et^{r}\Scon{\Al}rs \pi_s
     \end{equation}
 acting on the space
 $\Expower{}(\Lgd) \otimes \Expower{} (\Lg) \otimes \mathcal{F}(\Pmane) \otimes \Expower{}(\Lhd) \otimes \Expower{}(\Lh) $.
(Here as in Section~\ref{secBRST} we use $\pi_{\Al},\Et^a$ for elements of $\Lg$ and
$\Lgd$, while for the copy of $\Lh$ and its dual $\Lhd$ corresponding to the second factor in $\Sdpgh$ we use
$\Rh_{\Al}$ and $\Th^{\Al}$). If we define the function $\Ldiff{\Al}$ by
 \begin{equation}
\Ldiff{\Al}
 =\Pb{\Et^a \Mmap_a     - \Texthalf \Et^a \Et^b \Scon abc \pi_c}{\pi_{\Al}}
 =\Mmap_{\Al}-\Scon{\Al}{r}{s}\Et^r\pi_s
 \end{equation}
 the \Brst{} operator can be expressed in the form
\begin{equation}
    \Brstop =  \Et^a \Mmap_a
    - \Texthalf \Et^a \Et^b \Scon abc \pi_c -\Th^{\Al}(\Ldiff{\Al}-v_{\Al}).
     \end{equation}
This operator can be expressed as the sum of two commuting operators
$\BrstG=\Et^a \Mmap_a  - \Texthalf \Et^a \Et^b \Scon abc \pi_c$  and
$\BrstH=-\Th^{\Al}(\Ldiff{\Al}-v_{\Al})$.  (Each of these two parts
is canonical, independent of any choice of basis of $\Lg$ or $\Lh$.) The cohomology
corresponding to $\BrstH$ is acyclic, and explicitly solved by taking     functions which are
invariant under $\Rh_{\Al}$ and $\Ldiff{\Al}-v_{\Al}$. If we make the Kalkman transformation
 \cite{Kalkma}, that is, we conjugate by
$\Exp{-\Th^{\Al}\pi_{\Al}}$, we see that an equivalent cohomology is that of
 \begin{equation}\label{BSRTMODeq}
 \Brstop = \Et^a \Mmap_a - \Texthalf \Et^a \Et^b \Scon abc \pi_c + \Th^{\Al}v_{\Al}
 \end{equation}
where the auxiliary conditions are now $\Rh_{\Al}=\pi_{\Al}$ and
$\Ldiff{\Al}-v_{\Al}=0$.  (This transformation is the analogue of an extension due to Kalkman  \cite{Kalkma} of the Mathai-Quillen
isomorphism  \cite{MatQui} used in equivariant de Rham theory.)

This is one possible formulation of the \Brst{} operator of the theory. We will now use some
techniques from equivariant de Rham theory, which are also valid in this context, to give an
alternative formulation which allows quantization including a method for implementing the
auxiliary conditions by gauge fixing and corresponds to the \Brst{} operator for reducible
symmetries. Some terminology is required, which is summarised in Appendix~\ref{appWA},
following the book of Guillemin and Sternberg  \cite{GuiSte} where more details may be found.

The space $\Expower{}(\Lgd) \otimes \Expower{}(\Lg) \otimes \Fpman$ can be given the structure
of an $\Hstar$ algebra (Definition~\ref{HSTARdef}):  the $\tilde{\Lh}$ action is defined by
setting $i_{\Al}$ to act as Poisson bracket with $\pi_{\Al}$, $d$ to be the
\Brst{} operator $\BrstG$ and $L_{\Al}$ to act as Poisson bracket
with $\Ldiff{\Al}=\Mmap_{\Al}+\Scon{\Al}rs\Et^r\pi_s$, while  the
$H$ action is defined to be the adjoint action on $\Lg$, the
coadjoint action on $\Lgd$ and the original $H$ action on $\Pman$ with constraint map
$\Mmap_{\Al}$. It can further be given the structure of a $\Wstar$ module (Definition~\ref{HSTARdef})
if multiplication by the generator odd generator $\kappa^{\Al}$ of $W$ is given by Poisson
bracket with
$\Et^{\Al}$ and multiplication by the even generator
$u^{\Al}$ is given by Poisson bracket with $\Df \Et^{\Al}=-\Texthalf\Scon{a}{b}{\Al}\Et^a\Et^b$.

Another $\Wstar$ module $F$ may be defined by setting
 \begin{equation}
  F=\Fsmo(T^*(H))\otimes \Lambda(\Lhd) \otimes \Lambda(\Lh)=\Fsmo(T^*(H)\times\Real^{0,2l}) \, .
 \end{equation}
Taking coordinates $v_{\Al}, \Uvar^{\Al}, \theta^{\Al}$ and
$\Rh_{\Al}$ on $T^*(H)\times\Real^{0,2l}$ as before, the $\Hstar$ structure of $F$ is
defined by letting $i_{F\,\Al}$ act as Poisson bracket with
$-\Rh_{\Al}$, $L_{F\,\Al}$ with $-v_{\Al}$ and $\Df_F$
with $\sum_{\Al=1}^{l}\theta^{\Al}v_{\Al}$, while $H$ acts trivially on $\Lh$ and
$\Lhd$ and naturally on $T^*(H)$. The $W$ action on $F$ is defined by letting the generator
$\kappa^{\Al}$ of
$\Lhd$ act as Poisson bracket with  $\Th^{\Al}$, while the generators
$u^{\Al}$ of $W$ act by Poisson bracket with $w^{\Al}$.
The  basic cohomology of
 $\Brstop\otimes\One+\One\otimes\DiffF$ on
 $\Expower{}(\Lgd) \otimes \Expower{}(\Lg) \otimes \Fpman \otimes F$, that is, the cohomology
 on the subspace of $\Expower{}(\Lgd) \otimes \Expower{}(\Lg) \otimes \Fpman \otimes F$ whose
 elements  have zero Poisson bracket with $\mathcal{L}_{\Al}-v_{\Al}$ and with
 $\pi_{\Al}-\Rh_{\Al}$,
 is then the \Brst{}  cohomology of the doubly reduced phase space in the form (\ref{BSRTMODeq}). By
 Theorem~\ref{BCOHthe}, if
 we take  $E$ to be the Weil algebra $S(\Lhd) \otimes \Lambda(\Lhd)$ of $H$
 we see that an alternative form of the \Brst{} cohomology is the basic cohomology of
 \begin{equation}
    \Et^a \Mmap_a
    - \Texthalf \Et^a \Et^b \Scon abc \pi_c +u^{\Al}\rho_{\Al},
 \end{equation}
a form which suggests a close analogy with equivariant de Rham cohomology. (In this case the
basic conditions are zero Poisson bracket with $\mathcal{L}_{\Al}$ and with
 $\pi_{\Al}+\Rh_{\Al}$.) A gauge fixing procedure which implements these
basic conditions is constructed in  \cite{GFEC}.

A further possibility is the Cartan model, constructed by taking the Kalkman transformation as
before, which this time gives the \Brst{} operator in the form
 \begin{equation}
    \Et^a \Mmap_a
    - \Texthalf \Et^a \Et^b \Scon abc \pi_c +\Th^{\Al}\Ldiff{\Al}+u^{\Al}\rho_{\Al}-u^{\Al}\pi_{\Al},
 \end{equation}
 with basic condition $\Ldiff{\Al}=0, \Rh_{\Al}=0$ which is the same as the cohomology of
 $\Et^a \Mmap_a
    - \Texthalf \Et^a \Et^b \Scon abc \pi_c -u^{\Al}\pi_{\Al}$ on $\Ldiff{\Al}$ invariant elements of
$\Expower{}(\Lgd) \otimes \Expower{}(\Lg) \otimes \Fpman \otimes S(\Lhd)$.
  \section{Reducible symmetry}\label{secRSC}
In this section the notion of reducible symmetry as introduced by Batalin and G.A. Vilkovisky
\cite{BatVil,BatVil81} is related to the constrained systems whose reduced phase space and
\Brst{} quantization is considered in sections \ref{secRPS} and \ref{secMBRST}. The aim is, using
somewhat informal terminology based on the Lagrangian approach, to show how the particular
class of constrained system studied in this paper relates to the notion of reducible symmetry.
As already remarked, the work of Batalin and Vilovisky on systems with reducible symmetries
included the development of what has become known as
\Bv{} quantization, an extension of the \Brst{} technique, which gave a consistent functional
integral expression for the vacuum expectation value of theories with reducible constraints
(and has additionally led to a number of interesting developments in mathematics and physics,
involvong the master equation and odd symplectic manifolds, which are not considered in this
paper). These methods were further studied by Fisch, Henneaux, Stasheff and Teitelboim
\cite{FisHenStaTei} who gave an algebraic analysis of the
\Brst{} operator in terms of Koszul-Tate resolutions. A full account of these ideas may be
found in the book of Henneaux and Teitelboim  \cite{HenTei}, while an example of such a system
is considered in Section~\ref{secEX}.

For simplicity, so that the basic algebraic features are clear, we will restrict the
discussion at this stage to a quantum mechanical system where the symmetry of the system
corresponds to a finite dimensional group $G$ acting  on the fields $x^i(t), i=1, \dots,n$ of
the system. (The appendix shows how this may be extended to some infinite-dimensional group
actions.) Suppose that corresponding to a basis $\Set{\xi_a|a=1,\dots,m= \mathrm{dim}\, G}$ of
the Lie algebra $\Lg$ of $G$ the infinitesimal action of the group element
$1+\sum_{a=1}^{m}\Ep^a
\xi_a$ is
 \begin{equation}
    \delta_{\Ep} x^i = \Sum{a=1}{m}  \Ep^a(t) \Itrans{a}{i}(x)
 \end{equation}
where the $\Itrans{a}{i}$ satisfy
 \begin{equation}\label{eqGROUPPROP}
    \Itrans{a}{j}\Pbd{x^j}\Lrb{\Itrans{b}{i}} - \Itrans{b}{j}\Pbd{x^j}\Lrb{\Itrans{a}{i}}
    = \Sum{c=1}{m}\Scon{a}{b}{c} \Itrans{c}{i}
 \end{equation}
with $\Scon{a}{b}{c}$ the structure constants of $\Lg$ as before. If the action of a system is
 \begin{equation}
    S\Lrb{x(\cdot)} = \int \Df t \quad \Lagr\Lrb{x^i(t),\dot{x}^i(t)}
 \end{equation}
 and
 \begin{equation}
    \sum_{i=1}^{n}\Itrans{a}{i} \Dlagr{i} = 0 \Mboxq{for} a=1,\dots,m \, ,
 \end{equation}
 where $\Dlagr{i}= \Pbdt{\Lagr}{x^i}-\Dbd{t}\Lrb{\Pbdt{\Lagr}{\dot{x}^i}}$,
then the action is invariant under the action of $G$ and there are conserved Noether currents
 \begin{equation}
    \Curr_{a}=\Itrans{a}{i} \Pbdt{\Lagr}{\dot{x}^i}
 \end{equation}
which under Legendre transformation becomes constraints
 \begin{equation}
    \Mmap_{a}=\Itrans{a}{i} p_i  =0
 \end{equation}
where $p_i, i=1, \dots,n$ are the conjugate momenta of $x^i$. These constraints are first
class and obey the algebra
 \begin{equation}
    \Pb{\Mmap_a}{\Mmap_b}
    = \Sum{c=1}{m}\Scon{a}{b}{c} \Mmap_{c} \, .
 \end{equation}
As before we assume that the number $n$ of fields is at least as large as the dimension $m$ of
$G$.

If the $m$ vectors $\Itrans{a}{}$ are linearly independent for all $x$, or equivalently the
matrix $\Lrb{\Itrans{a}{i}}$ has rank $m$, the system is said to have an \emph{irreducible}
symmetry, and the \Brst{} procedure described in Section~\ref{secBRST} is applicable. A
\emph{reducible} symmetry occurs when the matrix has rank $m-l$, with $l>0$, in which case the
group property (\ref{eqGROUPPROP}) of the infinitesimal transformations will not in general be
satisfied. The concept of reducible symmetry was first identified by Batalin and G.A.
Vilkovisky \cite{BatVil,BatVil81}. In this paper we are concerned with  reducible systems for
which the infinitesimal transformations
$\Itrans{a}{i}$ are of the form
 \begin{equation}
       \Itrans{a}{i} =\Sum{b=1}{m}  \Reduc{a}{b}\Ghrep{b}{i}
 \end{equation}
where $\Lrb{\Reduc ab}$ is an $m\times m$ matrix of rank $m-l$ with $0<l<m$ and the elements
$\Ghrep{b}{i}$ have maximal rank $m$ and do satisfy the group property, that is,
  \begin{equation}
    \Ghrep{a}{j}\Pbd{x^j}\Lrb{\Ghrep{b}{i}} - \Ghrep{b}{j}\Pbd{x^j}\Lrb{\Ghrep{a}{i}}
    = \Sum{c=1}{m}\Scon{a}{b}{c} \Ghrep{c}{i} \, .
 \end{equation}
 (It may be conjectured that all reducible symmetries take this form.)
As a result there are $l$ non-trivial linear relations of the form
 \begin{equation}
    \Sum{a=1}{m} \La_{\Al}^{a} \Itrans{a}{}{} = 0 \qquad \Al=1,\dots, l
 \end{equation}
(where $\Itrans{a}{}$ denotes the vector $(\Itrans{a}{i})$) so that there are of course only
$m-l$ independent transformations, which will not in general form a Lie algebra. (Recent
discussions of Noether's second theorem
 \cite{FulLadSta,BasGiaManSar} consider related issues.) By a suitable choice of basis we can set
 \begin{equation}\label{eqREDCURR}
    \Itrans{\Al}{}{}=0, \quad \Al=1,\dots,l
    \Mboxq{and} \Itrans{r}{}{}=\Ghrep{r}{}{} \quad r=1+l,\dots,m \, .
 \end{equation}
On passing to the Hamiltonian formulation of the system, there will be
$m-l$ constraints
 \begin{eqnarray}
       \Mmap_{r} &\equiv& \Ghrep{r}{i}{}p_i = 0 \qquad r=1+l, \dots, m \, ,
 \end{eqnarray}
which in general will not form a first class system. Corresponding to the `missing' Noether
currents
$\Ghrep{\Al}{}{},\Al=1, \dots,l$
there are functions $\Mmap_{\Al}$ which are not constrained to be zero.  This leads to the
modified reduction process described in section \ref{secRPS}, involving an extended phase
space, and thence to the modified
\Brst{} quantization constructed in Section~\ref{secMBRST} which is equivalent  to the
\Brst{} operator obtained by the algebraic techniques of
\Bv{} quantization  \cite{BatVil,BatVil81,HenTei,FisHenStaTei}.

Chemla and Kalkman  \cite{CheKal} have shown that the \Brst{} operator for certain topological
theories corresponds to that for a system of reducible symmetries, using the transformation
$\Exp{\Th^{\Al}\pi_{\Al}}$ which we also use in this paper.  We have derived this result in a
more general context directly from the constraints of the system.  Outstanding questions
include analysing whether all reducible symmetries lead to constrained systems of this nature.
In the following section we give an example of the equivariant \Brst{} quantization of a
particular system.
\section{An example}\label{secEX}
As an example of the structures described in Sections~\ref{secRPS} and~\ref{secMBRST}, a
topological model will now be described. The setting of this model is an $n$-dimensional
Riemannian manifold $\Man$ with metric $g$ on which there is an isometric $\Uone$ action
generated by a Killing vector $X$. The classical action for this model is
 \begin{equation}
    S(x(.)) =
    \int_0^t  v \,x^* \Kof
 \end{equation}
where $x:[0,t] \to \Man$ is a path in $\Man$, $\Kof$ is the one form dual to $X$ via $g$ and
$v$ is a constant.  Using local coordinates $x^i, i=1, \dots, n$ on $\Man$ the action
takes the form
\begin{equation}
    S(x(.)) = \int_0^t  v \, X_i(x(t)) \dot{x}^i(t) \, \Df t \, ,
\end{equation}
where $X_i=g_{ij}X^j$ are the components of $\Kof$. The variational derivative of the
Lagrangian
$\Lagr\Lrb{x,\dot{x}}=v X_{i}(x)\dot{x}^i$ is
 \begin{equation}
    \Dlagr{i} =2v\Dc_j X_i\, \dot{x}^j = -2v\Dc_i X_j\, \dot{x}^j
 \end{equation}
(where $\Dc$ denotes covariant differentiation with respect to the Levi-Civita connection of
the metric $g$) so that the Lagrangian is invariant under infinitesimal transformations
$\delta x^i =
\Ep^i$ if $\Ep^i$  satisfies $\Ep^i X_i=0$. This gives $n$
reducible symmetries with one linear dependence relation. (It is here that the importance of
$X$ being a Killing vector first appears, since this ensures that
$\Textfrac12\Lrb{\partial_iX_j-\partial_j X_i} =\Dc_i X_j=-\Dc_j X_i$.)

Proceeding to the Euclidean time Hamiltonian formalism, the conjugate momentum to
$x^i$ is \begin{equation}
    p_i = i v X_i
\end{equation}
and we see that the system has $n$ first class constraints
$T_i \equiv p_i -i v X_i = 0$, which are of the general form (\ref{MCONSeq})
but in a geometrically natural basis rather than an $\Lh,\Lk$ basis. It is also necessary to
regard $v$ as a dynamical variable rather than a constant.  (Whether this is a general feature
of the constraints of a system with reducible symmetry is a question which needs further
exploration.) As expected the constraints of this system do not form a closed algebra; using
the standard symplectic form in the phase space $\Cotanm$ gives
 \begin{equation}
    \Pb{T_i}{T_j} = 2iv\Dc_i X_j\,.
 \end{equation}
This situation corresponds to that considered in Sections~\ref{secRPS}, \ref{secMBRST}
and~\ref{secRSC} (in the local version described in Appendix~\ref{appLA}) with
$\Gtil$ the diffeomorphism group of $\Man$ and $H$ the group $\Uone$ acting on $\Man$.
The groups $G$ which acts locally is then, as in Example~\ref{DIFFexa}, the $n$-dimensional
translation group
$\Real^n$ with constraint map $\Mmap_i =p_i$. To see that
the Lie algebra of $H$ has the required property, let $y$ be a point in $\Man$ where $X$
is not zero and $\Set{X} \cup\Set{\xi_r|r=2, \dots n}$ be a basis of the tangent space at
$y$ with each $\xi_r$ orthogonal to $X$.  Then, because $X$ is a Killing vector, it can
be shown that $\Comm{\xi_r}{X}$ is also orthogonal to $X$.  Thus if we identify $\Lk$ as
the span of $\Set{\xi_r|r=2,\dots,n}$ and $\Lh$ as the span of $\Set{X}$ we see that
$\Liebracc{\Lh}{\Lk}\subset\Lk$ as required.

In this case we have an $S^1$ extension, so that the modified \Brst{} procedure of
Section~\ref{secMBRST} gives as
\Brst{} operator
 \begin{equation}
     \Brstop = \Et^i p_i + u \Rh \,,
 \end{equation}
 with auxiliary conditions $\Rh=\pi$ and $\Ldiff{}=0$. (Since $H$ is
 one-dimensional we drop the index $\Al$.)
On quantization we obtain the differential in the Weil model of equivariant cohomology of
$\Man$ under the $\Uone$ action generated by $X$.  The full quantization of this model, using
the Kalkman form  \cite{Kalkma}, including gauge fixing, has been described in  \cite{GFEC}.
The model constructed in this section is equivalent to that constructed by Witten
 \cite{Witten82}, as can be shown by integrating out the $u,v,\Th$ and $\Rh$ variables or by
conversion to the Cartan model. It is also the same as that obtained by Chemla and Kalkman
 \cite{CheKal}. The derivation given in this section shows how the model can be understood as
the
\Brst{} quantization of a simple classical model, with the
conditions $\Rh=\pi$ and $\Ldiff{}=0$ emerging from the physics.
\appendix
\section{The local group action and the non-uniqueness of the constraints}\label{appLA}
In Section~\ref{secRPS} the standard approach to the reduced phase space of a constrained
Hamiltonian system was described, together with a modification for a more general set of
constraints than a standard first class set.

Even in the standard setting we have glossed over a difficulty which seems so far to have
received a rather incomplete treatment in the literature.  This relates to the non-uniqueness
of the constraint functions used to define the constraint submanifold, and hence the reduced
phase space. This non-uniqueness is a simple consequence of the fact that the constraints can
be multiplied by arbitrary nowhere-zero functions and still define the same constraint
submanifold, and satisfy a closed constraint algebra, although this will vary with the choice
of functions, and generally not form a finite-dimensional Lie Algebra. More fundamentally, it
relates to the fact that it is gauge or local symmetries which lead to the constrained
Hamiltonian systems discussed in this paper.

In this appendix we address these issues by describing the reduction process for the case
where the action of the finite group $G$ can only be identified locally, although there is a
group
$\Gtil$ (of infinite dimension) which acts globally on $\Pman$. It will emerge that essentially the
same construction of a reduced phase space can be made in this more general setting provide
that the $\Gtil$ action has properties which we will now define. (In the example below $G$ is
a diffeomorphism group, and we also have in mind the situation where $\Gtil$ is the group of
automorphisms of a fibre bundle, a so-called gauge group, but it is possible that even more
general situations may occur, and so the properties required are those that are essential.)
Suppose that $\Gtil$ is a Lie group which acts symplectically on
$\Pman$ and that there is an open cover  $\Set{U_{\Si}|\Si \in \LA}$
of $\Pman$  and, for each $\Si$ in $\LA$, a neighbourhood  $V_{\Si}$ of the identity of
$G$ such that  $V_{\Si}$ acts locally on $U_{\Si}$ in the following sense:
there is a map $ V_{\Si} \times U_{\Si} \to \Pman, (g,y) \to gy $ such that if $g,h$ and $gh$
are in
 $V_{\Si}$ and $y, hy$ are in $U_{\Si}$ then $(gh)y= g(hy)$. It is also required that the
local $G$ action is free, although the global $\Gtil$ action may have fixed points. Also
suppose that this local
$G$ action is compatible with the
$\Gtil$ action in that if
$\tilde{\Et}$ is an element of
$\Lgt$, the Lie algebra of $\Gtil$, then, for each
basis $\Set{\xi_a|a=1,\dots,m}$ of $\Lg$ (the Lie algebra of the finite group $G$) and each
$\Si \in \LA$ there exist $m$ functions
 $q_{\tilde{\Et}\, \Si}^a:U_{\Si} \to \Real, a=1,\dots,m$
such that for every $f$ in $\Fpman$
 \begin{equation}
 \Pvec{\tilde{\Et}} f\Resmap{U_\Si}
 = q_{\tilde{\Et}\, \Si}^a \, \Pvec{\xi_a}  f|_{U_\Si} \, .
 \end{equation}
(Here we again use the notation that $\Pvec{\tilde{\Et}}$ denotes the vector field on
$\Pman$ corresponding to the element $\tilde{\eta}$ of $\Lgt$, and so on.)

This group action is said to be Hamiltonian if both the global $\Gtil$ action and the local
$G$ action have constraint maps, denoted $\Jtil$ and $\Jsi$ respectively, with
 \begin{equation}
 \Jtil_{\tilde{\Et}}|_{U_\Si} = q^a_{\tilde{\Et} \Si} \, \Jsi_{a} \, .
 \end{equation}
The number of independent constraints is equal to the dimension of $G$ rather than
$\tilde{G}$.

An example of this structures will now be described.
 \begin{Exa}\label{DIFFexa}
Suppose that $\Pman$ is the cotangent bundle $\Cotanm$ of an $n$ dimensional manifold $\Man$,
$\Gtil$ is the diffeomorphism group of $\Man$ (which acts naturally on $\Cotanm$) and $G$ is
the
$n$-dimensional translation group $\Transg{n}$.  (As a manifold this group is simply
$\Real^n$.) We construct the open cover $\Set{U_{\Si}|\Si \in \LA}$ of $\Cotanm$ from an
open cover $\Set{W_{\Si}|\Si \in \LA}$ of $\Man$ by coordinate neighbourhoods, setting
$U_{\Si} = T^*W_{\Si}$. We then define the local action of $G=\Transg{n}$ by $(x^i,p_j)
\to (x^i+t^i,p_j)$ where $x^i,i=1,\dots,n$ are local coordinates on $W_{\Si}$, $(x^i,p_i)$
the corresponding local coordinates on $T^*W_{\Si}$ and $t^i,i=1, \dots,n$ is a sufficiently
small element of $\Real^n$. The local constraint maps for the $G$ action are
$\Mmap_i=p_i$. The Lie algebra of the diffeomorphism group of $\Man$ may be identified
with the set of vector fields on $\Man$.  If $Y$ is a vector field on $\Man$ with local
coordinate expression $Y=Y^i  \Pbd{x^i}$, then the global constraint map for $\Et$ is
 \begin{equation}
 \Jtil_{Y} = Y^i p_i \, .
 \end{equation}
 \end{Exa}

The two stage process leading to the reduced phase space can be carried out as before; in the
case where $\Pman$ has dimension $2n$ and the local group
$G$ has dimension
$m$ the reduced phase space will have dimension $2(n-m)$. In terms of constraints,
by proceeding to the larger group $\Gtil$, and allowing for the possibility of a local rather
than global action by the finite group $G$, we have explained the observed multiple
possibilities both for the set of constraints and for the algebra they form  \cite{HenTei}.

The modified reduced phase space corresponding to a subgroup $H$ of $G$ can also be handled in
this more general setting.  The requirement is a  finite-dimensional subgroup
$H$ of the global group which locally has the properties (\ref{HPROPeq}).
We can locally define the reduced phase space as before, except that there will be
singularities at fixed points of $H$.  Since (as will emerge from the example in
Section~\ref{secEX}) we can construct a non-singular \Brst{} operator even in this situation,
following the procedure which would be valid were there no fixed points, we will regard the
\Brst{} quantization scheme as the more fundamental object, and not pause to consider a fuller
definition of the reduced phase space in this context.
\section{The Weil algebra of $H$ and related constructions}\label{appWA}
In this appendix we gather some definitions and a theorem from equivariant de Rham theory,
using the book of Guillemin and Sternberg  \cite{GuiSte} where more details can be found.
 \begin{Def}\label{HSTARdef}
 Given an Abelian $l$-dimensional Lie group $H$ with Lie algebra $\Lh$,
 \begin{enumerate}\alphi
 \item  the super Lie algebra $\tilde{\Lh}$ is defined to be the
 algebra $\tilde{\Lh}=\Lh_{-1}\oplus\Lh_{0}\oplus\Lh_1$ where
  $\Lh_{-1}$ is an $l$-dimensional  vector space with basis $i_1,\dots,i_l$,
  $\Lh_{-1}$ is an $l$-dimensional  vector space with basis $L_1,\dots,L_l$ and
  $\Lh_1$ is one dimensional with basis $\Df$ and all Lie brackets are trivial except
  $\Liebracc{d}{i_{\Al}}=L_{\Al}$.
 \item A $\Hstar$ module is a super vector space $A$ together with a linear representation
 of $\phi$ of $H$ on $A$ and a homomorphism  of $\tilde{\Lh}\to\END{A}$

 which obey the consistency conditions:
 \begin{eqnarray}
    \frac{\Df}{\Df t} \phi(\Exp{t\xi})\Big|_{t=0} &=& L_{\xi} \,,\End
   \phi(a)L_{\xi} \phi(a^{-1}) &=& L_{\mathrm{Ad}_a\xi} \,, \End
   \phi(a)i_{\xi} \phi(a^{-1}) &=& i_{\mathrm{Ad}_a\xi}\,, \End
   \phi(a)\Df \phi(a^{-1}) &=& \Df \,.
 \end{eqnarray}
 \item The Weil algebra $W$ of the group $H$ is the algebra $S(\Lhd)\otimes\LA(\Lhd)$. The
 super algebra
 $\tilde{\Lh}$ acts on $W$ by superderivations with the only non-zero action on generators given by
 \begin{eqnarray}
       i_{\Al}(\One\otimes\kappa^{\Al}) &=& \One\otimes\One \End
  \Df (\One\otimes\kappa^{\Al}) &=& u^{\Al}\otimes \One
 \end{eqnarray}

where $\kappa^{\Al}$ are generators of
 $\LA(\Lhd)$ and $u^{\Al}$ are generators of $S(\Lhd)$.
 \item A $\Wstar$ module for the group $H$ is an $\Hstar$ module $E$  which is also a $W$
 module, with the map $W \otimes E \to E$ a morphism of $\Hstar$ modules.
 \item Corresponding algebra structures are defined when the vector spaces are algebras, $H$
 acts by automorphisms and $\tilde{\Lh}$ by superderivations.
 \end{enumerate}
 \end{Def}
A key theorem in equivariant de Rham theory is also valid in the form stated here.  It allows
the construction of alternative models of the equivariant \Brst{} cohomology. The proof may be
found in  \cite{GuiSte}.
 \begin{The}\label{BCOHthe}
Suppose that $E$ and $F$ are acyclic $\Wstar$ algebras and that $A$ is a $\Wstar$ module. Then
the cohomology of basic elements of $A\times E$ with respect to
 $\DiffA\otimes\One+\One\otimes\DiffE$
 is the same as that of basic elements of
 $A\times F$ with respect to
 $\DiffA\otimes\One+\One\otimes\DiffF$,
 where an element of $A\otimes E$ or $A\otimes F$ is
 said to be basic if it is annihilated by both
 $i_{\Al}\otimes\One+\One\otimes i_{\Al}$ and $L_{\Al}\otimes\One+\One\otimes L_{\Al}$.
 \end{The}

\noindent
\textbf{Acknowledgement} The author is grateful to J.D. Stasheff for comments on the first draft of this paper.
%
 %END OF TEXT **********************************
%

%
 \end{document}